\documentclass[%
reprint,
 amsmath,amssymb,
 aps,
floatfix
]{revtex4-1}
\usepackage{graphicx}
\usepackage{dcolumn}
\usepackage{bm}
\usepackage[utf8]{inputenc}
\usepackage{graphicx}
\usepackage{color}
   
\begin{document}
\title{Diversity waves in collapse-driven population dynamics}
\author{Sergei Maslov$^{1,2,3,*}$ and Kim Sneppen$^{4,+}$} 
\affiliation{
$^1$Department of Bioengineering, 
University of Illinois, Urbana-Champaign, IL  61801, USA\\
$^2$Carl R. Woese Institute for Genomic Biology, \\
University of Illinois, Urbana-Champaign, IL  61801, USA \\
$^3$Biological, Environmental and Climate Sciences Department, \\
Brookhaven National Laboratory, \\
Upton, NY 11973, USA \\
$^4$Center for Models of Life, 
Niels Bohr Institute, \\
University of Copenhagen, 2100 Copenhagen, Denmark\\
$^*$ssmaslov@gmail.com, 
$^+$ksneppen@gmail.com
}

\date{\today}

\begin{abstract}
\vskip 0.5 cm
\begin{center} 
{\bf Abstract}
\end{center}
Populations of species in ecosystems are often constrained by 
availability of resources within their environment.
In effect this means that a growth of one population, 
needs to be balanced by comparable reduction in populations 
of others. In neutral models of biodiversity all populations are 
assumed to change incrementally due to 
stochastic births and deaths of individuals.
Here we propose and model another
redistribution mechanism driven by abrupt and 
severe collapses of the entire population of a single species 
freeing up resources for the remaining ones. 
This mechanism may be relevant e.g. for 
communities of bacteria, with strain-specific 
collapses caused e.g. by invading bacteriophages, 
or for other ecosystems where infectious diseases 
play an important role.

The emergent dynamics of our system 
is cyclic ``diversity waves'' 
triggered by collapses of globally dominating populations.
The population diversity peaks at the 
beginning of each wave and exponentially decreases afterwards. 
Species abundances are characterized by a bimodal 
time-aggregated distribution with the lower peak formed by populations of 
recently collapsed or newly introduced species, while the upper peak 
- species that has not yet collapsed 
in the current wave. 
{\color{black} In most waves both upper and lower 
peaks are composed of several smaller peaks. 
This self-organized hierarchical peak structure 
has a long-term memory transmitted across several waves.
It gives rise to a scale-free tail of the 
time-aggregated 
population distribution with a universal 
exponent of 1.7. 
} 
We show that diversity wave dynamics is robust 
with respect to variations in the rules of our model 
such as diffusion between multiple environments, 
species-specific growth and extinction rates, 
and bet-hedging strategies.
\end{abstract}

\maketitle

\subsection*{Author Summary}
The rate of unlimited exponential growth is traditionally 
used to quantify fitness of species or success of organizations 
in biological and economic context respectively. 
However, even modest population growth 
quickly saturates any environment. Subsequent 
resource redistribution between the surviving populations is 
assumed to be driven by incremental changes due to 
stochastic births and deaths of individuals.
Here we propose and model another
redistribution mechanism driven by sudden and severe 
collapses of entire populations freeing up resources for the growth of 
others. The emergent property of this type of dynamics are 
cyclic ``diversity waves'' each triggered by a collapse of 
the globally dominating population.
Gradual extinctions of species within the current 
wave results in a scale-free time-aggregated 
distribution of populations of most abundant species. 
Our study offers insights to 
population dynamics of microbial communities with 
local collapses caused e.g. by invading bacteriophages.
It also provides a simplified dynamical description of 
market shares of companies competing in an economic sector 
with frequent rate of bankruptcy.
\subsection*{Introduction}
Mathematical description of many processes in biology and 
economics or finance assumes long-term exponential 
growth \cite{fisher1930,kelly} 
yet no real-life environment allows growth to continue 
forever \cite{malthus,verhulst}.
In biology any growing population eventually 
ends ups saturating the carrying capacity 
of its environment determined e.g. by nutrient 
availability. The same is true 
for economies where finite pool of new customers and/or natural
resources inevitably sets a limit on growth of companies.
Population dynamics in saturated environments is routinely described by 
neutral ``community drift'' models \cite{hubbell2001,woodcock2007} 
sometimes with addition of deterministic differences in efficiency of utilizing 
resources \cite{sloan2005}.

Here we introduce and model the saturated-state dynamics 
of populations exposed to episodic random collapses.
All species are assumed to share the same 
environment that ultimately sets the limit to their exponential 
growth. Examples of such systems include local 
populations of either a single or multiple 
biological species competing for the same 
nutrient, companies competing to increase their
market shares among a limited set of customers, etc. 
Specific case studies can be drawn from
microbial ecology where susceptible bacteria
are routinely decimated by exposure to bacteriophages 
(see e.g. \cite{middleboe2001,haerter2014} and references therein), 
or paleontological record, where 
entire taxonomic orders can be wiped out 
by sudden extinctions happening at a rate
independent of order size \cite{bornholdt2009}.

\subsection*{Model}
Population growth $P(t)$ of a single exponentially 
replicating species in a growth-limiting environment 
is traditionally described by Verhulst's \cite{verhulst} 
or logistic equation $dP/dt=\Omega \cdot P \cdot (1-P/P_{tot})$, 
where the carrying capacity of the environment 
$P_{tot}$ without loss of generality can be set to $1$.
In this paper we consider the collective dynamics of 
multiple populations competing for the same 
rate-limiting resource:
\begin{itemize}
\item 
Local populations are subject to episodic random collapses
or extinctions. The probability of an extinction
is assumed to be independent of the population size.
Immediately after each collapse the freed-up resources
lead to the {\color{black} transient} exponential population growth 
of all remaining populations $P_i$. The growth
stops once the global population 
$\sum_j P_j$ reaches the carrying capacity 
$P_{tot}=1$.
\item The environment is periodically reseeded 
with new species starting from the same 
small population size $\gamma \ll 1$ (measured 
in units of environment's carrying capacity).
\end{itemize}
We initially assume that the growth rates and 
collapse probabilities of all species are 
equal to each other. We also disregard the 
neutral drift \cite{hubbell2001} 
in sizes of individual populations during the time 
between subsequent collapses.
All these assumptions will be relaxed, simulated, 
and discussed later in the paper (see Supplementary 
Materials S1 Text, S1-S7 Figures).
The number of species in the steady state of the model 
is determined by the competition between the constant rate 
of introduction of new species and the overall rate of extinctions 
in the environment that is proportional to the number of 
species. To simplify our modeling we will consider a closely 
related variant of the model in which the number of species 
$N$ is kept strictly constant. In this case each extinction 
event is immediately followed by the introduction of a brand 
new species. We have verified that the two version of our 
model have very similar steady state properties.
The dynamics of the fixed-$N$ model is described by
\begin{equation}
dP_i/dt=\Omega \cdot P_i \cdot (1-\sum_j P_j) - \eta_i(t) \cdot P_i \quad ,
\label{logistic}
\end{equation}
where $\eta_i(t)$ is the random variable which is equal to zero 
for surviving populations 
and has a large positive value for populations 
undergoing an extinction/collapse.

To speed up our simulations we do not continuously  
calculate Eq. (\ref{logistic}) since most of the time the carrying capacity 
of the environment is saturated when local populations do not change.
Instead we simulate the model at {\color{black} discrete} time steps marked by extinction events.
At every time step a randomly selected local population goes extinct
and a brand new species with population $\gamma \ll 1$ is added to the environment.
We then instantaneously bring the system to its
the carrying capacity by multiplying all populations by 
the same factor. 

\subsection*{Results}

\begin{figure}[htp]
\centering
\includegraphics[angle=0,width=0.95\columnwidth]{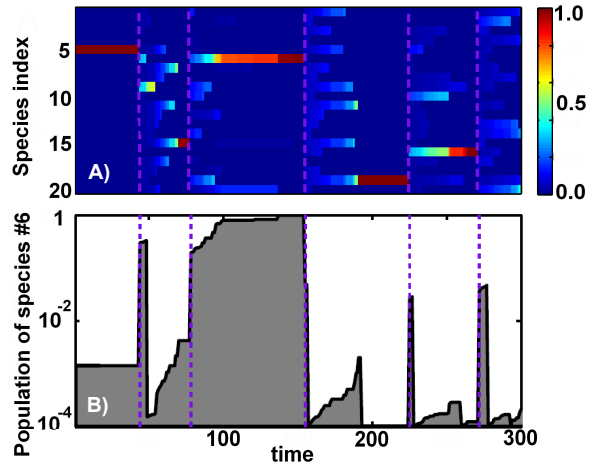}
\caption{{\bf Population dynamics.}
The simulated model has $N=20$ species and $\gamma=10^{-4}$.  
(A) Time-courses of populations of all species.
The color denotes population sizes 
(see the color scale on the right) with the dominating species 
visible as red horizontal bands. 
Note five diversity waves ending at purple dashed lines. 
Transitions between these waves were triggered
by extinctions of the dominating species \# 5, 15, 6, 19, 16 
correspondingly.
(B) The time-course of the species \# 6 with the logarithmic y-axis.
Note the pattern of intermittent periods of exponential growth fueled by 
local extinctions.
}
\label{fig1}
\end{figure}

In spite of its simplified description of the ecosystem 
disregarding pairwise interactions between species
our model has a rich population dynamics. 
Figure \ref{fig1}A shows the time-courses 
of populations in a system with  $N=20$ species and 
$\gamma=10^{-4}$. 
At certain times the carrying capacity of 
the environment is nearly exhausted by just one
dominant species with
$P_{max} \simeq 1$ visible as dark red stripes in Figure \ref{fig1}A. 
When such dominant species goes extinct 
a large fraction of the resources suddenly 
becomes available and consequently 
all other populations increase by a large ratio $1/(1-P_{max})$. 
This marks the end of one and the start of another 
diversity wave that initially is dominated by a large number of 
species with similar population sizes.
In the course of this new wave these species are eliminated one-by-one 
by random extinctions until only one dominant species is left 
standing. Its collapse ends the current and starts the new wave. 
In Figure \ref{fig1}A one can clearly distinguish about 
5 such waves terminated by the extinctions of dominant 
species \#5, 15, 6, 19, and 16 
correspondingly.

Figure \ref{fig1}B shows the time-course of just 
one local population of the species \#6 
on a logarithmic scale. 
Between time steps 100 and 150 
the population of this species nearly exhausts 
the carrying capacity of the environment.
Its local extinction at the time step 154 ended the third
diversity wave and started the fourth one.
Note somewhat erratic yet distinctly exponential 
growth of this species happening on 
the slow timescale set by the frequency of extinctions. 
This growth should not be confused 
with exponential re-population of recently collapsed 
environments that happens much faster 
(a small fraction of one time step).

\begin{figure}[htp]
\centering
\includegraphics[angle=270,width=0.95\columnwidth]{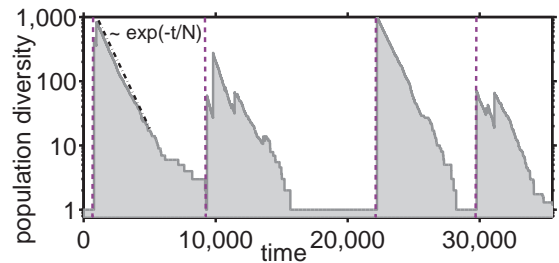}
\caption{{\bf Diversity dynamics.} The grey shaded area shows the the 
time course of the population 
diversity $D=1/\sum_i P_i^2$ in our model 
with $N=1000$ and
$\gamma=10^{-12}$. 
Purple dashed lines mark the beginnings of 
diversity waves when a collapse of the dominant species with
$P_{max} \simeq 1$ leads to an abrupt increase in population diversity 
from $\sim 1$ to $\sim N$. The diversity subsequently 
decreases $\propto \exp(-t/N)$ (dash-dotted line)
}
\label{fig2}
\end{figure}

Figure \ref{fig2} follows the population diversity (grey shaded area)
defined as $D(t)=1/\sum_{i=1}^N P_i(t)^2$ as a function of time in a 
system of size $N=1000$. In general $D$ can vary between 
$\sim 1$ when one abundant species 
dominates the environment and $N$ when 
all species are equally abundant.
The diversity is inversely proportional to the 
largest population $P_{max}(t)=\max_i P_i(t)$. 
The diversity waves (purple dashed arrows in Figure \ref{fig2})
are initiated when a dominating species collapses.
As a consequence, at the start of each wave 
the diversity abruptly increases from 
$\sim 1$ to a substantial fraction of the maximal possible 
diversity $N$.
After this initial burst triggered by the global 
redistribution of biomass, the diversity 
exponentially declines 
as $\exp(-t/N)$ (the dot-dashed line in Fig.\ref{fig2}), 
driven by ongoing extinctions of 
individual populations.
The typical duration, $t_{wave}$ of a diversity wave 
is equal to the time required for all of the 
species present at the start of the wave to 
go extinct one-by-one. Thus it is determined 
by $N \cdot \exp(-t_{wave}/N) \sim 1$ or
\begin{equation}
t_{wave}\sim N \cdot \log_{e}N \qquad .
\end{equation}

\begin{figure}[htp]
\begin{center}
\includegraphics[angle=270,width=0.95\columnwidth]{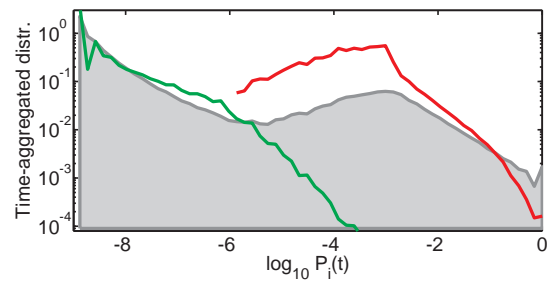}\\
\end{center}
\caption{{\bf Time-aggregated population size distribution.}
{\color{black} The grey shaded area shows the time-aggregated} distribution
of population sizes in our model with $\gamma=10^{-9}$ 
and $N=1000$ collected over 20 millions collapses.  The green and red 
lines show the population size distributions collected, respectively, 
at the very end of each wave and at the very beginning of 
the next wave correspondingly as described in the text.
Note that they roughly correspond to two peaks of the 
{\color{black} time-aggregated} distribution.
}
\label{fig3}
\end{figure}

Figure \ref{fig3} shows the {\color{black} time-aggregated} distribution 
of population sizes for $\gamma=10^{-9}$ and $N=1000$
(the grey shaded area). This logarithmically-binned distribution
defined by $\pi(P)=d\mathrm {Prob}(P_i(t)>P)/d \log_{10} P$ 
were collected over 20 million individual collapses
(time-steps in our model). Thus, a {\color{black} time-aggregated} distribution
is rather different from a typical ``snapshot'' of the system
at a particular moment in time characterized by 
between $1$ and $N$ of highly abundant species in 
the current diversity wave. The 
{\color{black} time-aggregated} distribution is bimodal 
with clearly separable peaks corresponding 
to two population subgroups. 
The upper peak consists of the species that have not 
yet been eliminated in the current wave. 

To better understand the dynamics of the system in 
Figure \ref{fig3} we also show the 
distribution of populations sizes 
at the very end of each diversity wave (green line) 
and at the beginning of the next wave (red line). 
That is to say, 
for the green curve 
we take a snapshot of all populations
immediately after the dominant species with 
$P_{max} > 1-1/N$ was eliminated,
but before 
the available biomass  was redistributed among all species.
At those special moments, happening only once every 
$t_{wave}$ time steps,  
most population sizes are between $\gamma$ and $\gamma \cdot N$ while 
a small fraction reaches all the way up to $\sim 1/N$. 
During the rapid growth phase immediately after 
our snapshot was taken, all populations grow by the same 
factor $1/(1-P_{max}) \simeq N$ thereby moving all 
of them to the upper peak of the 
{\color{black} time-aggregated} distribution 
thereby starting the new wave. The red curve corresponds to 
the snapshot of all populations immediately after this 
rescaling took place. It shows that 
at the very beginning of the new wave 
local populations 
are broadly distributed between 
$\sim N\cdot \gamma$ and $1$ with a peak 
around $1/N$.

\begin{figure}[htp]
\begin{center}
\includegraphics[angle=270,width=0.95\columnwidth]{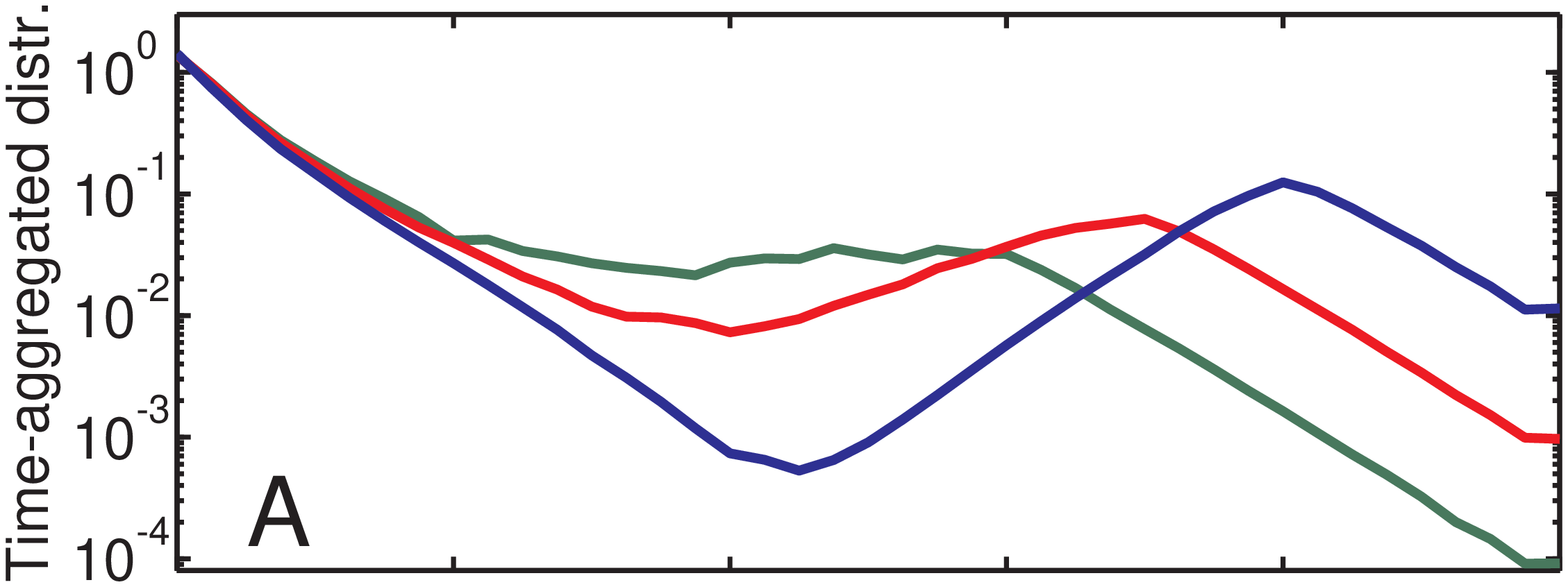}\\
\includegraphics[angle=270,width=0.95\columnwidth]{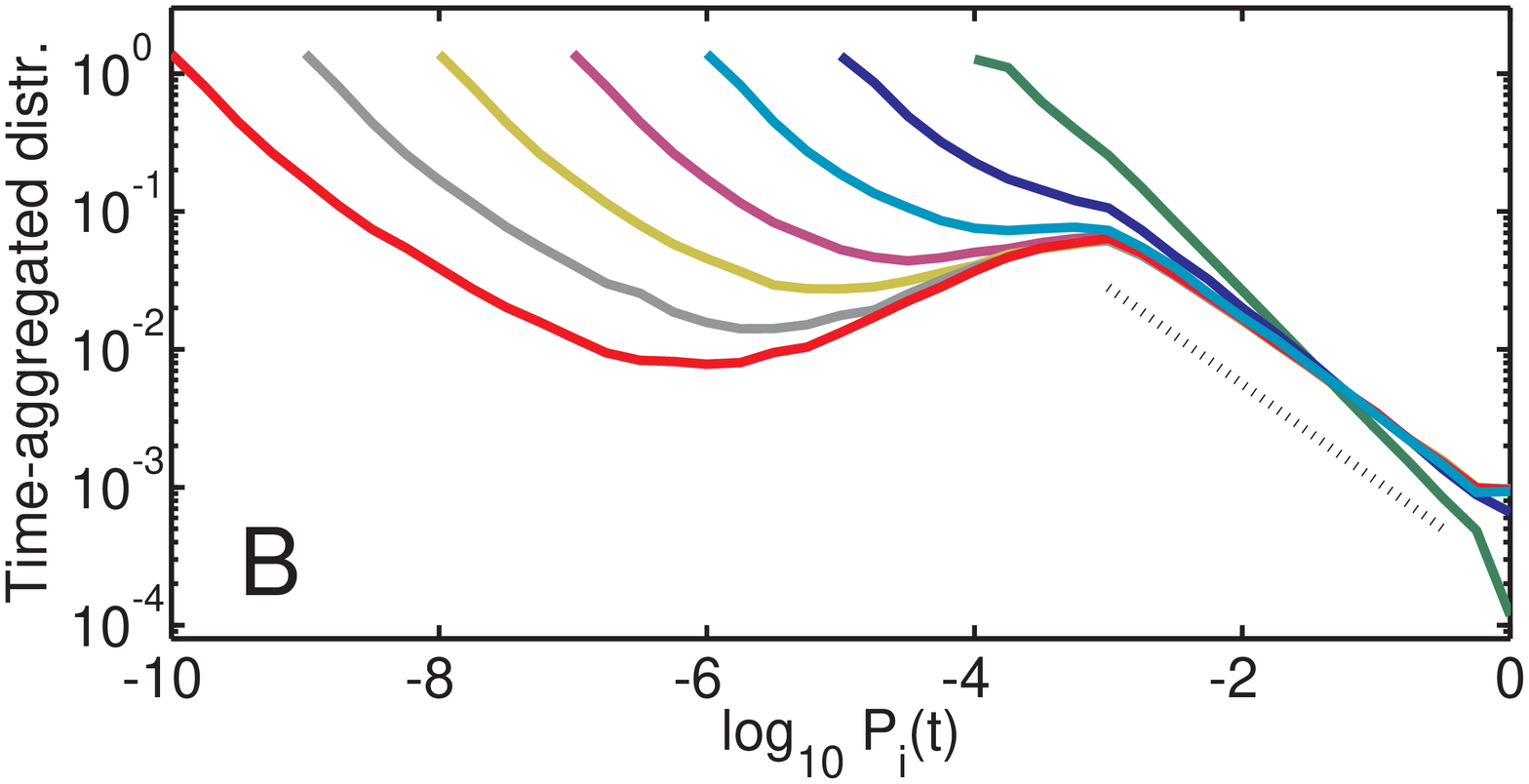}\\
\end{center}
\caption{{\bf  Time-aggregated distributions for different values of $N$ and $\gamma$.}
{\color{black} Time-aggregated} distributions of population sizes 
in our model with 
A) $\gamma=10^{-10}$ and $N=100$ (blue), $N=1000$  (red), and 
$N=10,000$ (green). 
B) $N=1000$ and varying $\gamma$ ranging between $10^{-4}$ (green) 
to $10^{-10}$ (red) 
in ten-fold decrements. 
Note the emergence of a nearly universal scale-free 
tail of the distribution fitted with $\tau \simeq 1.7$ (dashed line).
}
\label{fig4}
\end{figure}

Figure \ref{fig4}A shows {\color{black} time-aggregated} distributions 
of population sizes for $\gamma=10^{-10}$ and different 
values of $N$ 
{\color{black}
ranging between
}
 $100$ ad $10,000$ 
while  Figure \ref{fig4}B shows 
{\color{black} time-aggregated} distributions 
with $N=1000$ and for a wide range of $\gamma$. 
One can see that for $\gamma<0.01/N$,
the tail of the distribution for most abundant populations 
between $1/N$ and $1$ is well fitted by a power law 
$\pi(P) \propto 1/P^{\tau-1} \simeq 1/P^{0.7}$ (dashed line in Figure \ref{fig4}B)
corresponding to the power law distribution of population sizes on the 
linear scale $d \mathrm{Prob}(P_i(t)>P) /d P \sim 1/P^{\tau} \simeq 1/P^{1.7}$.
Overall Figures \ref{fig4}A,B demonstrate that
the exponent $\tau$ for different values of $\gamma$ and $N$ 
is remarkably universal. Indeed, for a range of parameters 
where the upper and the lower peaks of $\pi(P)$ are clearly separated, 
$\tau$ approaches a universal value $\tau=1.7$.

{\color{black}
An insight into the origins 
of the scale-free tail of the distribution 
of population sizes is gained by considering a simplified 
version of our model in which at the start of each wave 
the populations of all species are artificially set to be equal to each other
resulting in the maximal diversity. 
We further assume that $\gamma \ll 1/N$.
The passage of time $t$ elapsed since the beginning of the current wave, 
leads to a decrease in the number of surviving species 
$N_{surv}(t)=N \exp(-t/N)$, 
which all have the same population size $P=1/N_{surv}(t)$ 
jointly filling up the carrying capacity of the 
environment.  Above we ignore a negligible fraction ($\sim \gamma$) 
of the total population of the lower peak  species.
The time-aggregated probability for a species to have a population size
$P_i >P=1/N_{surv}(t)$ is naturally given by $N_{surv}(t)/N \propto 1/P$
and thus 
\begin{eqnarray*}
\mathrm{Prob}(P_i >P) &\propto &\frac{1}{P}  \Rightarrow  \\
\mathrm{Prob}(P_i =P) & = &\frac{d\mathrm{Prob}(P_i >P)}{dP} \propto \frac{1}{P^2}
\end{eqnarray*}
The exponent $\tau=2$ predicted by this simplified model 
is realized in our actual model 
for moderately high $\gamma \sim 0.1$, whereas
smaller values of $\gamma$
give rise to a different universal exponent $\tau \simeq 1.7$.  
The decrease of the exponent $\tau$ from 2 to 1.7 in our original model 
is the result of unequal population sizes at the beginning of a new wave.
In fact, we verified numerically that  $\tau=2$ is recovered if at the start of each wave 
one equilibrates all species abundances to $1/N$.
The first section of the S1 Text in supplementary materials 
provides additional details on how 
the reduced population diversity $D=1/\sum P_i^2<N$ 
at the start of population waves affects the exponent $\tau$.

Two panels in Figure \ref{fig4new} illustrate the difference between the 
simplified (panel A) and the real (panel B) models. In both versions of the model the 
average jump in the logarithm of surviving populations grows exponentially 
with time $t$ elapsed since the start of the current wave: 
$-\log(1-P_{collapsed}(t)) \simeq exp(t/N)/N$.
However, unlike the simplified model, the population distribution in 
our real model has a rich hierarchical structure with multiple 
sub-peaks in some waves (color bands in Figure \ref{fig4new}B).
Remarkably this multi-modal distribution reappears
in subsequent waves, implying that the memory about the hierarchical structure 
in the upper part of the distribution is transmitted 
to emerging populations in the lower part 
with sizes starting at $\gamma$. At the start of the next 
wave these same populations 
will move to the upper part of the distribution thereby 
transmitting the history
across waves.
Colors of symbols in Figure \ref{fig4new} 
illustrate the origin of multiple peaks. 
Indeed, populations in each of these peaks 
were born during the previous wave under similar conditions
(the number of substantial populations) as described in the caption.
Thus, the broadening of time-aggregated population distribution 
in our model compared to its history-free version is a simple manifestation 
of a complex interplay between "upstairs" and "downstairs" 
subpopulations transmitting memory across waves.
\begin{figure}[htp]
\begin{center}
\includegraphics[angle=270,width=1\columnwidth]{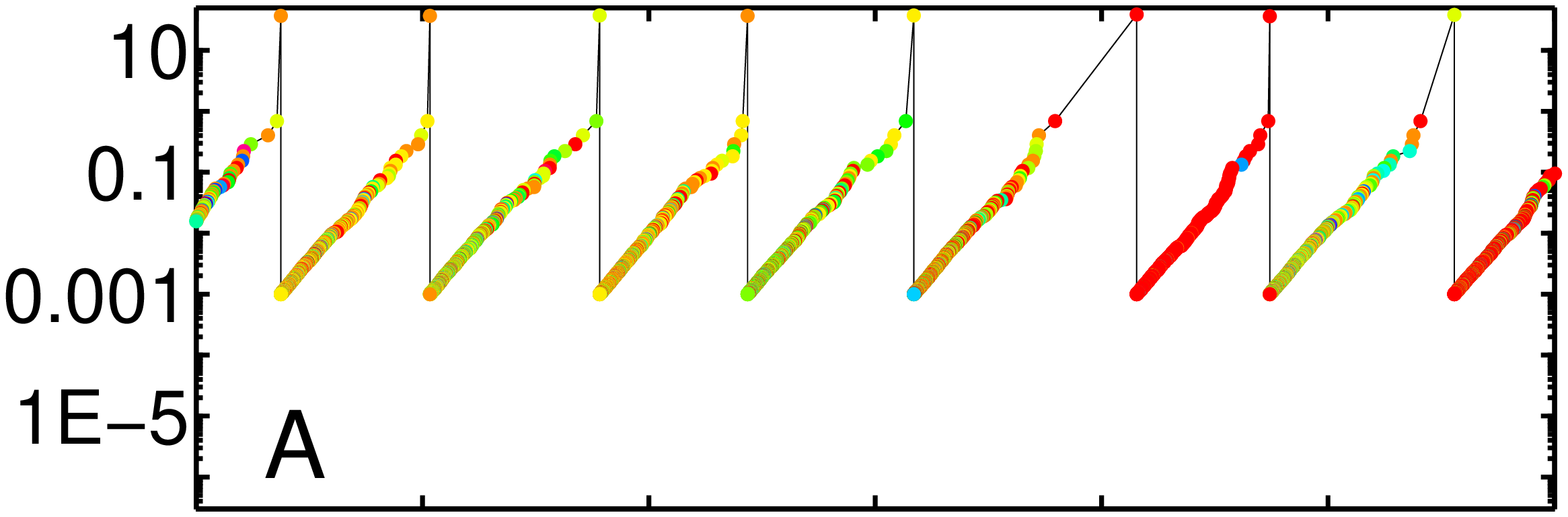}\\
\vspace{-1.0cm}
\includegraphics[angle=270,width=1\columnwidth]{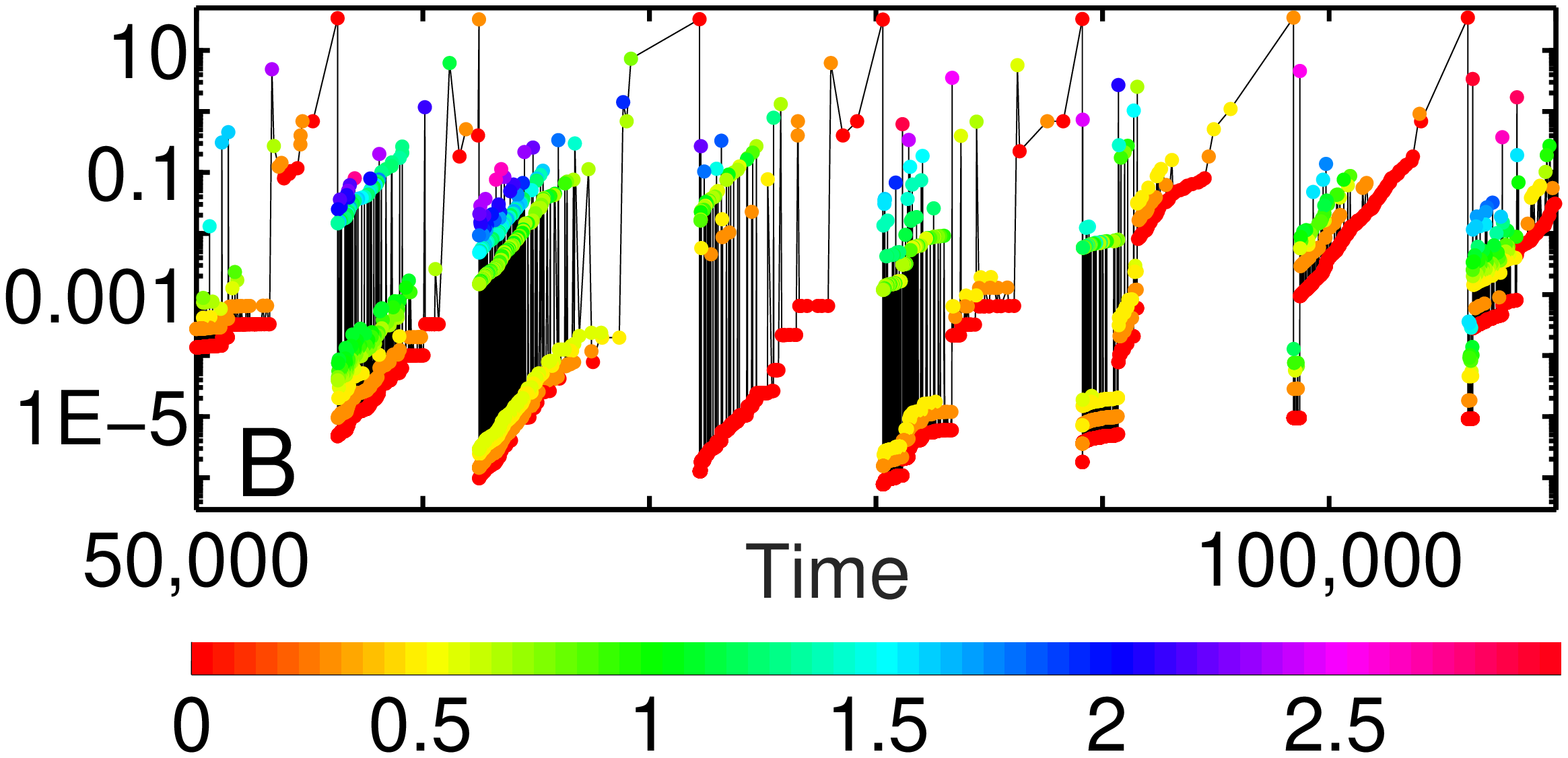}\\
\end{center}
\caption{{\bf  Memory of  population size distribution is preserved across several diversity waves. } 
Time course of jumps $-\log[1-P_{collapsed}(t)]$ 
in the logarithm of surviving populations following a collapse of a 
substantial population $P_{collapsed}(t)>10^{-10}$ in 
A) the simplified model in which at the start of each wave 
all populations are set equal to each other;
B) our basic model. Both were simulated at $N=1000$ and 
$\gamma=10^{-20}$. 
Note that our basic model, unlike its simplified counterpart, preserves 
memory of population sizes distribution across several subsequent diversity waves.
This is manifested in similar fractal structure of jumps sizes in waves 
\#2-6 shown in panel B). Colors of symbols represent the 
$log10$ of the number of substantial populations during the 
the previous wave, when a given population originated at the small 
size $\gamma$.  Thus red dots mark populations originated at  the very end of the 
previous wave, while yellow dots - those originated when there were 
two large populations left in the previous wave. Finally, green, blue, 
and purple dots refer to older populations in the previous wave.
}
\label{fig4new}
\end{figure}

The population distributions in both upper and lower peaks in our 
model are described by the same exponent $\tau$. 
This similarity reflects the fact that individual populations in the lower peak
are exposed to the same multiplicative growth as the ones in the upper 
peak.  Finally, the intermediate region of the distribution connecting two peaks
is shaped by rescaling of all populations in the lower peak 
as they are moved up at the beginning of a new diversity wave.
When peaks are well separated (as e.g. for low values of $\gamma$ in Figure \ref{fig4}) 
the slope of the logarithmic distribution in this region
has almost exactly the same value $\tau-1=0.7$ and the opposite
sign to the slopes in both the upper and the lower peaks.
}

\subsection*{Discussion}
In this paper we explore the population dynamics in saturated 
environments driven exclusively by near-complete collapses of 
sub-populations of competing species. This type of dynamics 
strongly contrasts the  gradual changes implied in for example 
the ``community drift'' neutral models \cite{hubbell2001} 
in ecology, or for the most part incremental random walks 
of stock values of individual companies. Conversely, 
collapse-driven dynamics represents a sudden and usually large change 
of system composition. In 
ecology such collapses may 
be caused e.g. by invading predators or diseases
, whereas in the economy,  
companies of any size routinely go bankrupt e.g. through excessive 
debt amplifying the effects of external perturbations.  

{\color{black} 
First, let us consider biological systems. 
One of the predictions of our model is 
a multimodal logarithmic distribution of population sizes.
Indeed, while the time-aggregated distribution is 
bimodal with distinct upper and lower peaks, populations within 
any given diversity wave cluster together in several smaller 
peaks persisting over several waves (color stripes in 
Figure \ref{fig4new}B).
This overall finding is supported by a growing body of 
literature \cite{gray2005,dornelas2008,vergnon2012,matthews2014} 
where multi-modal Species Abundance Distributions (SAD) in 
real ecosystems were reported for plants, 
birds, arthropods \cite{matthews2014}, 
marine organisms including single cells,  corals \cite{dornelas2008}, 
nematodes, fishes, entire seafloor communities  \cite{gray2005},  
and even extinct brachiopods
\cite{olszewski2004}. Like in our model, the empirical SADs
range over many orders of magnitude with a 
noticeable depletion (or several depletions) 
at intermediate scales. The magnitude of this dip is 
usually somewhat less than predicted by our basic model but is consistent
with several of its variants described below. This includes the model 
variant \#1 inspired by the neutral theory of biodiversity 
\cite{hubbell2001} thought to apply to a variety of ecosystems including 
microbial communities \cite{sloan2005,woodcock2007} (see S1 Figure 
in supplementary materials). 

Needless to say, our model is not unique in generating 
multimodal distributions (see e.g. \cite{vergnon2012} for other examples).
Conversely, some of the variants of our model
give rise to interesting population dynamics including diversity waves even 
without any depletion in the middle of the log-binned SAD. 
We argue that a more reliable characterization of underlying dynamical 
processes can be obtained from time-series data. 
First, all systems capable of diversity waves 
are described by rapid large changes in populations of 
individual species.
}
Such sudden, population-scale shifts can occur e.g. due to 
introduced diseases or the disappearance of keystone 
species \cite{paine1969,cohn1998} thereby changing 
the composition of the entire food-web. On the microbial 
scale, sudden invasion of a new bacteriophage may 
lead to multiple orders of magnitude reduction in the 
population of susceptible bacteria 
\cite{levin1977,middleboe2001}, potentially leading to 
their complete local extinction \cite{haerter2014}. 
Phage-driven collapses do not spare bacteria with large 
populations and may even be biased towards such 
{\color{black}
as postulated in the Kill-the-Winner (KtW) hypothesis \cite{thingstad1997}. 
The magnitude and characteristic timescales of population 
changes in microbial ecosystems is still being actively 
discussed in the literature. 
While Ref. \cite{campbell2011} reports that over half of all bacterial 
species in marine environments varied between abundant and 
rare over a three-year period, other studies \cite{rodriguez-brito2010}
found more modest variability at the level of species or genera over 
weeks to months period.
However, everyone seem to agree on dramatic and rapid 
(often on the scale of days \cite{castberg2001}) 
population shifts at the level of individual bacterial strains 
\cite{middleboe2001,middelboe2009,rodriguez-brito2010}
caused by phage predation \cite{castberg2001}. 
Except for interchangeable gene cassettes (metagenomic islands) 
responsible for either phage recognition cites or 
alternatively resistance to phages 
\cite{rodriguez-valera2009} 
these strains routinely have very similar genomes and thus may have 
near identical growth rates. 
Hence, they are capable of coexistence in the saturated 
state implicitly assumed in our model. 

Extinctions and collapses in our model are assumed to be 
caused exclusively by exogenous effects such as natural 
catastrophes or predation by external species not sharing 
the carrying capacity of our environment. Real-life ecosystems 
can also collapse due to endogenous effects, i. e. internal 
interactions between species. Such intrinsic collapse mechanisms 
were the sole focus of earlier models by us and others 
(see e.g. \cite{bak1993,sole1996,haerter2014}).
}

On vastly longer, geological timescales,  
the collapse-driven dynamics of our model 
resembles that of species extinctions and subsequent radiations 
in the paleontological record \cite{raup,vanvalen}. 
One example is the recolonization by mammals of a number of ecological 
niches vacated by dinosaurs after the end-Cretaceous mass extinction
{\color{black}
thought to be preceded by a gradual depletion of diversity among dinosaurs
who were finally wiped out by a singular catastrophic event \cite{sloan1986}. 
}
Interestingly, the extinction rate of taxonomic orders appears to be 
independent of their size quantified by the number of 
genera they contains \cite{bornholdt2009}, which is also one of 
the assumptions of our basic model of collapse-driven dynamics.  

The second area of applications of our model is to describe 
company dynamics in economics. The size or the market share 
of a publicly traded company reflected in its stock price is 
well approximated by a random walk with (usually) small 
incremental steps \cite{bachelier1900}. However, as in the case 
of ecosystems, this smooth and gradual drift does not account 
for dramatic rapid changes such as bankruptcies or market crashes. 
In case of companies the main driver of sudden changes is 
their debt \cite{fisher1933}. When the intrinsic value of a 
company is much smaller than its debt, even small changes in 
its economical environment can make it insolvent not sparing 
even the biggest companies from bankruptcies (think of Enron 
and Lehman Brothers). Empirically, the year-to-year volatility 
of company's market share varies with its size $S$, 
$\Delta S/S \propto S^{-0.2}$ \cite{econo_hurst}.

Abundance distributions in our original model 
and many of its variants are characterized by 
a power-law tail with an exponent $\tau$ close to 
$2$. 
This is in approximate agreement with the empirical data on 
abundance distributions of bacteria in soil samples 
\cite{bacterial_zipf}, marine viruses (phages) \cite{viral_zipf}. 

In the economics literature, the distributions 
of company sizes  
\cite{filiasi2014}, as well as those of wealth of individuals 
\cite{klass2004} are known to have similar scale-free tails.
Recent data for companies \cite{filiasi2014} 
and personal wealth \cite{klass2004} suggest
$1/P^{1.8}$ tail of the former distribution and $1/P^{2.3}$ tails of the latter one. 
Traditionally, scale-free tails in these distributions were explained by 
either stochastic multiplicative processes pushed down against 
the lower wall (the minimal population or company size, or welfare 
support for low income individuals) 
\cite{levy1996,levy1997,sornette}, by variants of rich-get-richer dynamics 
\cite{simon1955}, or in terms of Self-Organized Criticality 
\cite{bak1987,bak1993}. The emphasis of the latter type of models
on large system-wide events (avalanches \cite{bak1987,bak1993} or collapses
\cite{newman1996}) and on separation of timescales is similar in spirit to 
the collapse-driven dynamics in our models. 
{\color{black} A potentially important socio-economic implication 
of our model is that during each wave contingent 
sub-peaks in the ``upstairs'' part of the distribution 
are imprinted on the ``downstairs" part and thereby can be repeated 
in the new wave following the ``revolution''.
}

Needless to say, our models were simplified 
in order to compare and contrast the collapse-driven 
dynamics to other mathematical descriptions of competition 
in saturated environments. The S1 Text in  
supplementary materials describes several 
variants of our basic model that in addition to population 
collapses include the following elements:
\begin{enumerate}
\item {\color{black} ``Neutral drift model''} 
assumes changes of population sizes during 
time intervals between collapses as described in Ref. \cite{hubbell2001}.
In this model in addition to collapses a population of size $P_i$ randomly drifts up and down $\Delta P_i \propto \pm \sqrt{P_i(1-P_i)}$. The 
resulting diversity waves and {\color{black} time-aggregated} distributions can be found 
in the supplementary S1 Figure.
\item {\color{black} ``Exponential fluctuations model"} is another variant of the neutral scenario
where the population sizes
between collapses undergo slow multiplicative adjustments
$\Delta P_i \propto \pm \Omega_i P_i$ restricted by the overall 
carrying capacity of the environment. Details and the resulting 
{\color{black} time-aggregated} distribution can be found in the 
supplementary S2 Figure.
\item {\color{black}``Interconnected environments model" is  
another neutral variant of our basic rules} in which 
spatially separated sub-populations of the same species are
 competing with each other for the same nutrient. Sub-populations are connected 
by the diffusion, that is much slower than the diffusion of shared 
nutrient. In this model a collapsed sub-population is replenished by a small number 
$\gamma$ of individuals diffusing from other environments, 
see the supplementary S3 Figure.
\item {\color{black} "Kill-the-Winner (KtW) model"} 
where collapse probability $c$ 
systematically increase with the population size as 
suggested by the studies of phage-bacteria ecosystems
\cite{thingstad1997}. In this particular case 
the diversity dynamics 
and the scale-free tail of the population distribution 
becomes sensitive to the extent that the large populations 
are disfavoured by collapse. When the collapse probability
is proportional to population size, 
one obtain a flat  distribution where numbers of species 
in each decade of population sizes are equal to each other, see 
the supplementary S4 Figure.
\item {\color{black} "Kill-the-Looser (KtL) model"}, 
where collapse probability $c$ 
systematically decreases with the population size $P$ as 
$c(P) \sim P^{-0.2}$ as suggested by the empirical studies of 
company dynamics \cite{econo_hurst}.
As seen in the supplementary S5 Figure the diversity dynamics 
and the scale-free tail of the population distribution 
are both remarkably robust with respect to introduction 
of size-dependent collapse rate.
\item {\color{black} "Fitness model" in which} each of the species
{\color{black} has} its own 
growth rate $\Omega_i$ during rapid re-population phase and 
its own collapse probability $c_i$. 
The supplementary S6 Figure show that the overall shape of 
the {\color{black} time-aggregated} distribution is similar 
to that in our basic model, whereas its lower panel illustrate the
interplay between population size and 
the the two variables that define the species' fitness.
\item {\color{black} ``Resilience model" as a variant of 
the above fitness scenario}, in which 
collapsing species do not necessarily go into extinction.
Instead, each species is assigned its own ``survivor
ratio'' $\gamma_i$ defined by the population drop following a 
collapse: $P_i \to \gamma_i P_i$. As in the previous variant 
each of the species is also characterized by its own 
growth rate $\Omega_i$. The supplementary S7 Figure 
shows that for intermediate populations 
the {\color{black} time-aggregated} distribution is 
described by a power law scaling.
Compared to the basic model it has 
a broader scaling regime and larger likelihood to
have most of the ``biomass" collected in one species.
\end{enumerate}
Captions to supplementary S1-S7 Figures provide more detailed description 
of model dynamics in each of these cases. Overall, the `
simulations of the variants of our basic model  
described above preserve the general patterns of 
collapse-driven dynamics such as  
diversity wave dynamics, and a broad 
{\color{black} time-aggregated} distribution of population sizes
with scale-free tail for the most abundant species.

\begin{figure}[htp]
\vspace{0.5cm}
\centering
\includegraphics[angle=270,width=0.95\columnwidth]{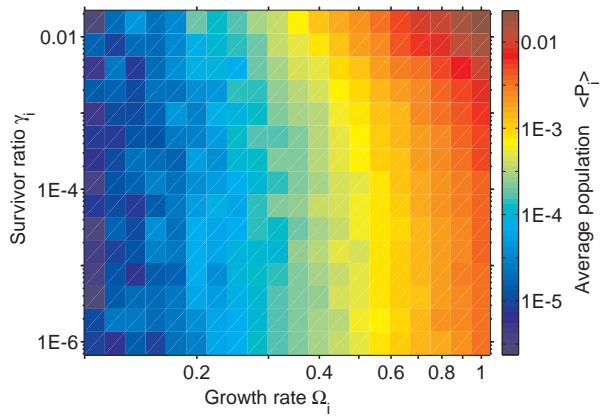}
\caption{{\bf  Average population vs species' 
properties in the ``Fitness model'' variant \#6.}
Time-averaged population of a species (see color scale
on the right) plotted as a function of its re-population
growth rate $\Omega_i$ (x-axis) and
population drop after collapses $\gamma_i$ (y-axis). 
in a variant of our model with fitness differences 
between species. Note that the population 
increase with both $\Omega_i$ and $\gamma_i$.
Populations and fitness parameters 
of $N=1000$ species were taken from 50 million
snapshots of the model.
}
\label{fig5}
\end{figure}

The classic definition of the fitness of a species in terms 
of its long-term exponential growth rate 
{\color{black} \cite{fisher1958}} is clearly 
inappropriate for our model.  Indeed, the 
long-term growth rate of each of our species 
is exactly zero.  {\color{black} One must keep in mind though that 
fitness is a very flexible term and has been defined in 
in a variety of ways \cite{orr2009} reflecting (among other things) 
different timescales of growth and evolution 
\cite{bak1993}, and relative emphasis on population 
dynamics vs. risk minimization \cite{bergstrom2004}.
An appropriate way to} quantify species' 
success in a steady state system like ours is in terms of 
their time-averaged population size $\langle P_i(t) \rangle_t$.  

In the last two variants of our basic model we add 
fitness-related parameters to each of the species: its short-term
exponential growth rate $\Omega_i$ (model 6 and 7), 
its propensity to large population collapses quantified by their 
frequency $c_i$ (model 6), and the severity of collapses 
quantified by surviving fraction $\gamma_i$ of the population (model 7). 
Fig. 6 shows the average
population size $\langle P_i(t)\rangle_t$ 
as function of $\Omega_i$ and $\gamma_i$ in the model 6.
As one can naively expects species with 
larger short-term growth rates and larger surviving 
ratios also tend to have substantially larger populations
(the red area in the upper right corner of Fig. 6).

While in our models the probability $c_i$ and severity 
$1/\gamma_i$ of collapses are assumed to be independent 
of growth rate $\Omega_i$ in reality they are 
often oppositely correlated.
Indeed, in biology much as in 
economics/finance fast growth 
usually comes at the cost of 
fragility and exposure to downturns
forcing species to optimize trade-offs. 

Some environmental conditions 
favor the fast growth even at the cost 
of a higher risk of collapse, 
whereas other could call for sacrificing 
growth to minimize probability or severity of collapses.
Species in frequently collapsing environments 
considered in our study are known to employ 
bet-hedging strategies 
\cite{kelly,zhang,bergstrom2004,ms_kelly2015} 
to optimize their long-term growth rate.
This is obtained by splitting their 
populations into ``growth-loving'' and 
``risk-averse'' phenotypes 
\cite{bergstrom2004,kussell2005,ms_kelly2015}.
One example of this type of bet-hedging 
is provided by persistor sub-populations of some bacterial species 
consisting of $\gamma \sim 10^{-4}$ of the overall population 
\cite{baleban2004,gerdes2013} increasing to as much as 
$\gamma=10^{-2}$ 
in saturated conditions (S. Semsey, private communications).
A bet-hedging strategy with persistor sub-population $10^{-2}$ 
somewhat reduces the overall growth rate (only by 1\%) 
while dramatically reducing the severity of collapses caused 
by antibiotics. 
From Fig. 6 one infers that this is indeed a good trade-off.

{\color{black}
In this study we presented a general modeling framework for systems
driven by redistribution of rate-limiting resources freed up by
episodic catastrophes. In spite of their simplicity the population
dynamics in such systems happens on at least {\bf four hierarchical
timescales}. At the shortest timescale the populations grow
exponentially repopulating resources vacated during a catastrophic
extinction event. This exponential growth results in a steady state 
at which the system is poised exactly at the carrying capacity 
of the environment. At even longer timescales the system is described 
in terms of diversity waves that are the main focus of this study.
These waves are an emergent dynamical property of the system 
in which  population of surviving species grow exponentially while 
species diversity decays.  Remarkably the information about the 
``upstairs''and ``downstairs'' population peaks survives the ``revolution''
at the end of each wave. This memory gives rise to the final, longest 
timescale in our system correlating several consecutive waves. All of 
this complexity is already present in our basic one-parameter model. 
In spite of its simplicity this model and its variants provide the foundation 
for future studies of collapse-driven dynamics in ecosystems, 
market economies, and social structures. 
}

\begin{center}
\subsection*{References:}
\end{center}

\begin{small}

\end{small}

\renewcommand{\theequation}{S\arabic{equation}}
\setcounter{equation}{0}
\renewcommand{\thefigure}{S\arabic{figure}}
\setcounter{figure}{0}
\cleardoublepage 
\newpage
\section*{Supplementary Materials}
\subsection*{Fokker-Planck equation for the basic model}
{\color{black}
Let's consider a version of our model with a very low value of 
$\gamma$ to ensure the complete separation 
between the upper and lower peaks.  
The multiplicative dynamics of surviving populations in the upper peak 
is described by the following elementary step:
$$P_i(t)=\frac{P_i(t-1)}{1-P_{j(t)}(t)}$$. Here $P_j(t)(t)$ 
is the population at time step 
$t$ of the species $j(t)$ that went extinct at this time step.
It can be also easily integrated for all times since the beginning 
of the current wave at time step $1$:
\begin{equation}
P_i(t)=\frac{P_i(1)}{1-\sum_{t'=1}^t P_{j(t')}(1)} \qquad .
\label{eq.s1}
\end{equation} 
Indeed, $\sum_{t'=1}^tP_{j(t')}(1)$ is the total initial (at time step $1$) 
populations of all species that went extinct by the time step $t$. 
Hence $1-\sum_{t'=1}^tP_{j(t')}(1)$
- is the total initial population of all surviving species used to 
normalize their initial populations 
to give their populations at the time step $t$ (population ratios of surviving species are preserved in our basic model).
Taking the logarithm of both sides of Eq. \ref{eq.s1} and approximating $-\log (1-\sum_{t'=1}^tP_{j(t')}(1)) \simeq 
\sum_{t'=1}^tP_{j(t')}(1)$, which holds as long as the system is still far away from the end of the wave ($\sum_{t'=1}^tP_{j(t')}(1) \ll1$, 
 one gets:
\begin{equation}
\log P_i(t)=\log P_i(1) + \sum_{t'=1}^tP_{j(t')}(1) \qquad .
\label{eq.s2}
\end{equation} 
The stochastic dynamics within a single wave can thus be described 
by the following equation:
\begin{equation}
\frac{d \log(P_i(t))}{dt}=P_{j(t)}(1) \qquad .
\label{eq.s2p}
\end{equation}

The exponent $\tau$ of the population size distribution in our model is determined by the balance between 
the noisy multiplicative population dynamics and the exponential loss of surviving species due to collapses. 
It can be approximated by the following Fokker-Plank-like equation:
\begin{equation}
\frac{\partial \pi }{\partial t}=-\rho \pi -\mu \frac{\partial \pi }{\partial \log P}+ \sigma \frac{{{\partial }^{2}}\pi }{{{(\partial \log P)}^{2}}}+\text{episodic source terms }
\label{eq.fp}
\end{equation}
Here $\pi (\log P,t)$ is the time-dependent population abundance distribution in the upper peak, $\rho $ - the loss term due to population collapses, $\mu $ - the logarithmic drift velocity and $\sigma$ - is the logarithmic dispersion, which is totally absent in the simplified model
where all populations start at the same size. 

The Eqs.\ref{eq.s2}-\ref{eq.s2p} allow us to derive the parameters of the Fokker-Plank equation in terms of the distribution of 
population sizes $P_i(1)$ at the start of the wave. Indeed, early in the wave one has $\rho=1/N$, $\mu=\langle P_i(1) \rangle_i=1/N$ and 
$\sigma=\langle P_i(1)^2 \rangle_i-\langle P_i(1) \rangle_i^2=(1/N)\cdot (1/D(1)-1/N)$. Note an unusual connection between the population diversity at the start of a wave $D(1)$, and the diffusion coefficient $\sigma$ in the Fokker-Plank equation. 

The stationary solution for the time-aggregate distribution $\int{\pi }(\log P,t)dt$ has an exponential tail $\exp (-(\tau -1)\log P)$. It corresponds to 
the power law tail of the species population distribution $\propto 
{{P}^{-\tau}}$. The exponent $\tau $ is defined by one of the two 
solutions to the quadratic equation 
\begin{equation} 
0=-\frac{1}{N}+\frac{1}{N}(\tau -1)+\sigma{{(\tau -1)}^{2}} \qquad .
\label{eq.quadratic}
\end{equation}
In the version of the model, where all populations at the start of the wave are equal to each other, sizes of surviving populations 
increase deterministically as $\exp(t/N)/N$ (see main text for the derivation) and thus have zero dispersion: $\sigma=0$.
Hence in this simplified version the exponent $\tau =1+\rho/\mu=2$ is determined by balancing only the first two terms of this equation. 
 
We have numerically verified that the decrease of the exponent 
from $\tau=2$ in the simplified model 
down to  $\tau =1.7$ in our original model is driven entirely 
by noise (unequal population sizes) resulting in a finite value of $\sigma$.
A non-zero value of $\sigma$ in the Eq. \ref{eq.quadratic} results in $\tau<2$.  
For example, if the populations at the start of each wave had a Poisson distribution 
so that $\sigma=\mu=1/N$, 
the exponent $\tau=(1+\sqrt{5})/2 \simeq 1.62$ would have been defined 
by the solution of the golden mean equation $0=-1+(\tau-1)+(\tau-1)^2$.
While currently we have no first-principles argument allowing us to derive the value 
of $\sigma$ in our basic model, the result from a Poisson distribution is not 
too far from the empirically observed exponent $\tau =1.7$

\subsection*{Model variants}
To test the robustness of our basic model with respect to rule 
changes we considered the following seven variants:
\begin{enumerate}
\item {\bf ``Neutral drift model''}. 

This variant extends our basic model 
by adding to our standard model the random 
neutral drift of population sizes 
(Hubbell SP (2001) The unified neutral 
theory of biodiversity and biogeography (MPB-32), 
Princeton University Press)
between subsequent collapse events.   
To simulate this random drift, at every time step 
the population of each species changes up or down 
as prescribed by dispersion of binominal distribution:
$P_i\rightarrow P_i\pm \sqrt{r \cdot P_i (1-P_i)}$, where 
$r$ is the parameter quantifying the magnitude of fluctuations 
proportional to the inverse of the total population size and 
the square of the birth/death rate.
After drift changes were applied to all populations 
we rescale them back to their 
carrying capacity $\sum P_i=1$. 
This is followed by a collapse event as in the standard model.
Fig. \ref{figS1} 
illustrates typical time courses 
of the diversity $D(t)=1/\sum P_i(t)^2$ and 
time-aggregated species abundance distributions 
in this model variant for three values of $r$ 
and compares them to our basic model.
\begin{figure}[htp]
\centering
\includegraphics[angle=0,width=0.85\columnwidth]{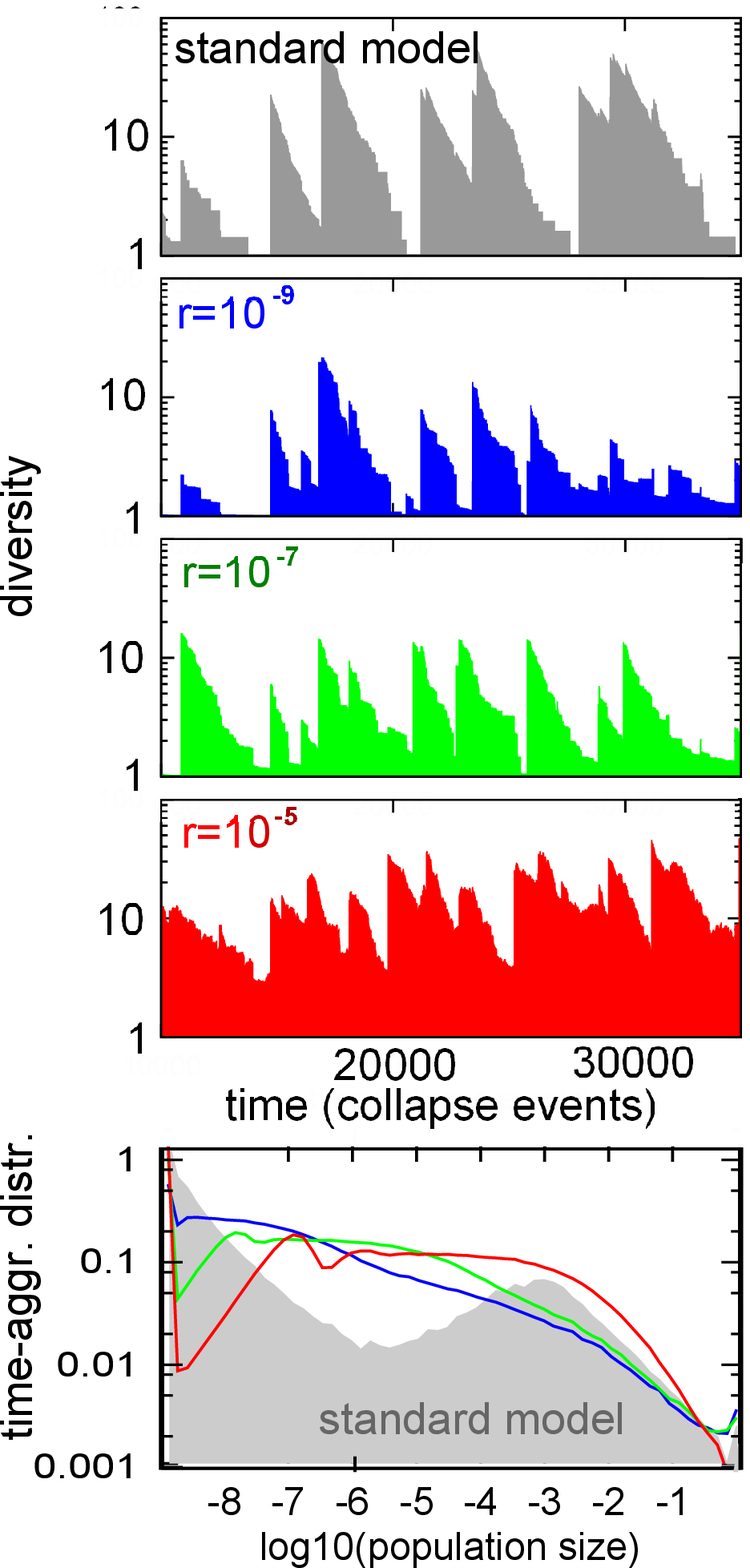}
\caption{\it {\bf ``Neutral drift model''}. 
This variant extends our basic model 
with $N=1000$ and collapse ratio
$\gamma=10^{-9}$ by adding the 
neutral drift at rate $r$ taking place 
between subsequent collapse events in 
our standard model: $P_i\rightarrow P_i\pm \sqrt{r \cdot P_i (1-P_i)}$.   
The lower panel shows the {\color{black} time-aggregated} distributions in 
our system simulated for $10^6$ collapse events.
The grey shaded area refers to our basic, unmodified model, i.e.
to the $r=0$ case, while three color lines correspond to 
$r=10^{-9}$ (blue), $r=10^{-7}$ (green), and $r=10^{-5}$ (red).
The upper four panels illustrate typical time courses 
of the diversity $D(t)=1/\sum P_i(t)^2$ in our basic model and 
for three values of the $r$ color-coded as in the lower panel.
}
\label{figS1}
\end{figure}

\item {\bf ``Exponential fluctuations model''}, 
where populations are exposed
to random exponential shifts 
between successive collapse events.
In this version of the model between successive collapse events all populations exponentially shift thus adjusting the way 
carrying capacity is divided between the. This model is similar to the 
"Neutral random drift" model \#1 above except that changes are proportional to $P_i$ and not to $\sqrt{P_i}$. The population of each species is characterized by its own exponential rate $G_i(t)$ 
given by $n \cdot rand_i(t)$, where $n$ quantifies the overall rate of the redistribution, while 
$rand_i(t)$ - a random number uniformly distributed between 0 and 1 - represents species- and time- specific shifts.
We assume that differences in growth rates between species 
are not constant but instead fluctuate on the timescale 
when a single species collapses. 
Hence, after each collapse event we reset the exponential growth we 
randomly reset the 
rates $G_i(t)$ for all species (and not just of the collapsed one). 
In this version of the model we also take into account the stochastic 
nature of the time interval $\tau_r(t)$ between two successive 
collapse events, which is randomly chosen from the exponential 
distribution with mean value equal to 1. 
Thus between two successive collapse events each species population
changes as $P_i\rightarrow P_i \cdot
e^{G_i(t)\tau_r (t)}$ and subsequently 
rescaled to the carrying capacity of the environment: $\sum P_i=1$. 
As in our basic model, at every time step one randomly selected population $i$ 
collapses and is reset to 
$P_i\rightarrow \gamma$ while all other populations
are rescaled to fill up the carrying capacity: $\sum P_i=1$. 
Fig. \ref{figS2} shows the simulations of this model for several 
value of $n$  compared to our basic ($n=0$) model.
\begin{figure}[htp]
\centering
\includegraphics[angle=0,width=0.85\columnwidth]{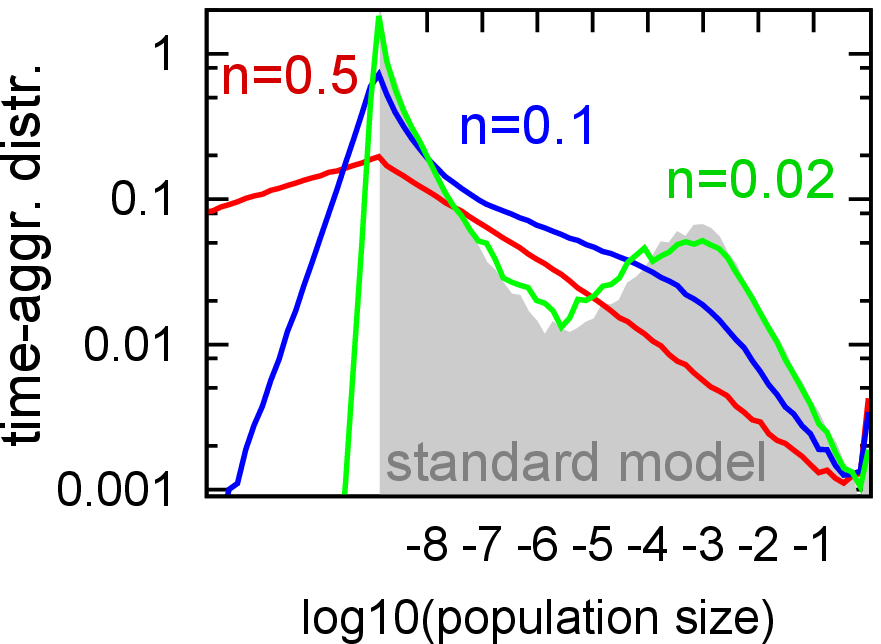}
\caption{\it {\bf ``Exponential fluctuations model''} 
with $N=1000$, $\gamma=10^{-9}$, and $n=0.02$ (green), $n=0.1$ (blue), 
$n=0.5$ (red) system simulated for $10^6$ collapse events.
The grey shaded area shows the time-aggregated population 
distribution in our basic model, corresponding to the $n=0$ 
limit.
}
\label{figS2}
\end{figure}

\item {\bf ``Interconnected environments model''} with diffusion. 
In this version of the model there is a {\bf single}
species distributed between $N$ local environments.  
As in our basic model at every time step 
one local population $i$ is selected for collapse and reset to 
$P_i \rightarrow 0$ after which the populations are normalized 
back to the carrying capacity. The diffusion takes a fraction 
$\gamma$ of the total population of $1$ and distributes it equally 
between all local environments: 
$P_i \rightarrow P_i(1-\gamma)+\gamma/N$.
Note, that here we implicitly assume that populations in 
all of these environments share the same carrying capacity. This 
is the case when diffusion rate of the rate-limiting 
nutrient is much faster then that of populations themselves. 
The main difference of this model from earlier variants is that 
populations in the lower peak with $P_i \ll 1/N$ grow approximately 
linearly in time (as opposed to exponentially in other 
versions of the model). The rate of this linear growth is the same for 
all species and is equal to $\gamma/N$. The exponential growth is
restored for populations that are larger than average.
Fig. S3 shows the time-aggregated distribution of local 
population sizes in this model variant.
\begin{figure}[htp]
\centering
\includegraphics[angle=0,width=0.85\columnwidth]{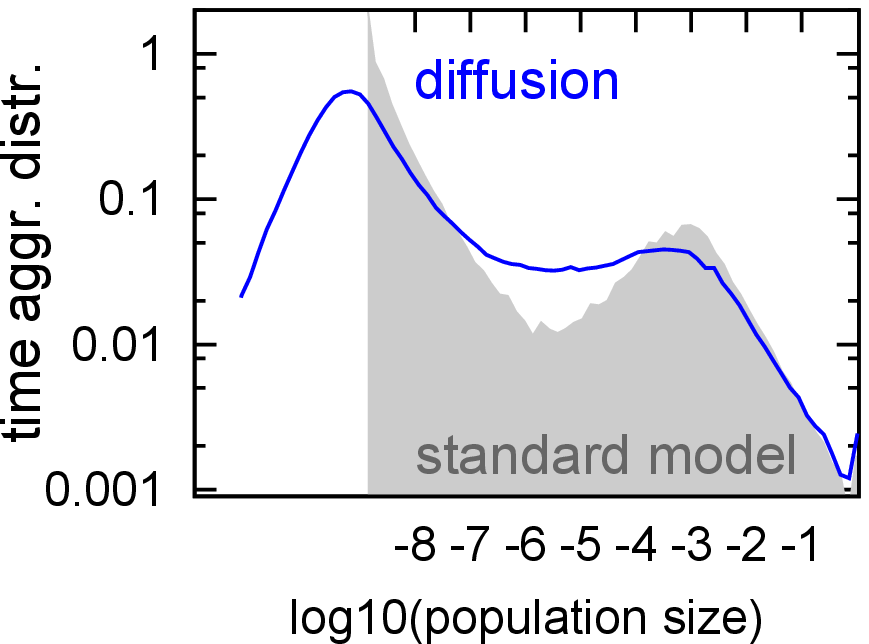}
\caption{\it {\bf `` Interconnected environments model''}. Figure show time-aggregated species abundance distributions 
in the model with $N=1000$ environments connected by diffusion 
of strength $\gamma=10^{-9}$ simulated over  
$10^6$ collapse events.
The standard model with the same 
parameters is shown as the grey shaded area. 
}
\label{figS3}
\end{figure}

\item {\bf ``Kill-the-Winner (KtW) model''}.
For bacterial populations
the direction of the trend (if any) of collapse probability 
with population size is currently unknown. In fact one can 
plausible make a case for increasing of the probability 
of collapse with population size due to larger populations
making easier to find and overall 
and more attractive targets for phages.
In microbiology preferential targeting of large bacterial 
populations by virulent phages is 
known as "Kill-the-Winner" (KtW) hypothesis
(Thingstad TF and Lignell R (1997)
Aquatic Microbial Ecology 13:19-27).  
Here we simulate the version of our basic
model where the collapse probability systematically 
increases with population size.  
At each time step we select a random population to collapse 
with probability $\propto P_i^{\sigma}$.
As before the collapsing population is reset to
$P_i \rightarrow \gamma$ and
all populations are subsequently 
grown with equal exponential 
rates to complete saturation: $\sum P_i=1$. 
Fig. S4 examines time-aggregated population distributions in 
KtV model variant for different values of $\sigma$.
Whereas small and moderate $\sigma$ preserve 
diversity wave dynamics, the $\sigma=1$ version 
does not exhibit diversity waves and predicts a 
population size distribution distributions, $dP/ds\propto 1/s$,
or $dP/d\log(s)=constant$ (equal number of species in each decade of 
population sizes).
\begin{figure}[htp]
\centering
\includegraphics[angle=270,width=0.85\columnwidth]{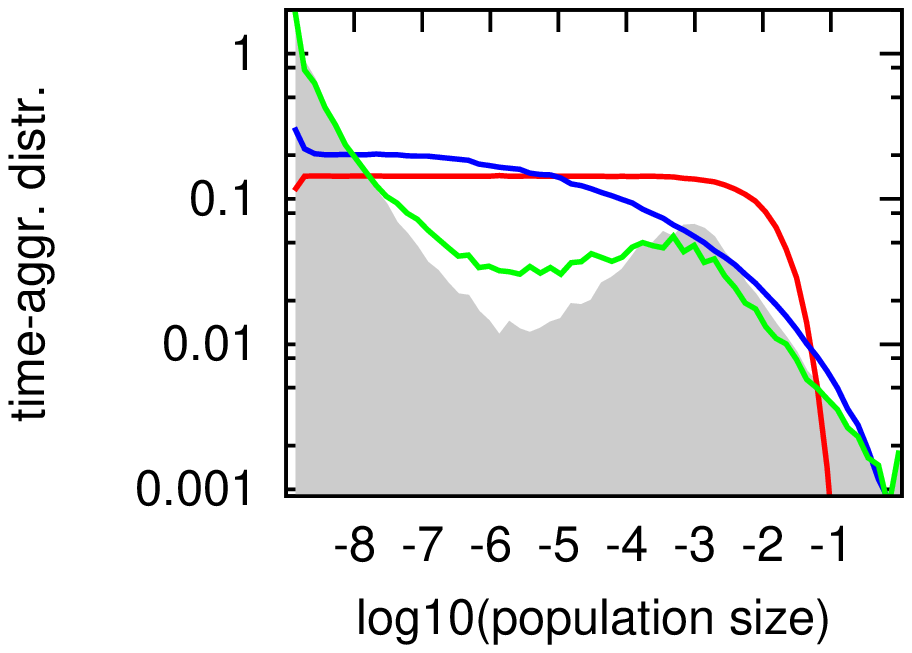}
\caption{\it {\bf ``Kill-the-Winner (KtW) model''} in which 
larger populations are preferentially targeted for collapse:
$c_i \propto P_i^{\sigma}$. Different colors correspond to 
time-aggregated SADs in the model with $N=1000$, $\gamma=10^{-9}$, and 
$\sigma=0.01$ (green), $0.2$ (blue), and  $\sigma=1.0$ (red) simulated 
over $5 \cdot 10^6$ collapse events. The grey shaded area refers to 
time-aggregated population distribution in our basic, 
unmodified model with the same $N$ and $\gamma$.
}
\label{figS4}
\end{figure}

\item {\bf ``Kill-the-looser (KtL) model''} in which a collapse probability 
declines with population size in a power-law fashion.  
At each time step one select a random population to collapse 
with probability $\propto P_i^{-0.2}$ where $P_i$
is the current population size of species $i$.
In economics this corresponds to an intuitively plausible
notion that larger companies are less likely to go bankrupt 
than smaller ones. Empirically, this trend is described by 
a power law with the exponent -0.2 (Nunes Amaral LA, Buldyrev SV, 
Havlin S, Leschhorn H, Maass P, Salinger MA, et al. 
(1997) Journal de 
Physique I. 7: 621-633.) 
Notice the emergence of the lower peak distribution distribution above 
$\gamma$ that is much more narrow than in our standard model. 
This makes sense as in the course of each wave small populations tend 
to collapse over and over. These repeated collapses don't drive other populations 
up by much and thus their only consequence is clustering of small populations 
close to the very bottom of the lower peak distribution at and above $\gamma$. 
When the dominant upper peak population finally collapses 
all small populations are rescaled up to form a narrow distribution around 
$1/N$. This is very similar to our simplified memory-free model described 
in the main text and shown in Fig. 4A. 
The new wave starts with very high 
diversity $D(t) \simeq N$ which is subsequently reduced  
with time as $D(t)=N_{surv}(t) \simeq N \exp(-t/N)$. 
Here we ignore a relatively small $N^{0.2}$-fold decline in collapse frequency over the range of 
population sizes between $1/N$ and $1$. 
Within the same approximation 
each surviving population grows as $P(t) \propto \exp(t/N)$.
The time-averaged distribution of populations thereby
approaches the scaling regime described by:
\begin{eqnarray}
\nonumber
\mathrm{Prob}(P_i(t) >P) & \sim & \frac{1}{P} 
\Rightarrow 
\\
\frac{d\mathrm{Prob}(P_i(t) >P)}{dP} & \sim & \frac{1}{P^2}
\end{eqnarray}
In reality the scaling exponent of the tail is around $1.8$. 
It is the same as in our standard model but for a 
different reason.
Indeed, taking into account that lifetime of a population before
collapse scales as $1/P^{-0.2} =P^{0.2}$ one gets
$\mathrm{Prob}(P_i(t)=P) \sim \frac{P^{0.2}}{P^2}=\frac{1}{P^{1.8}}$.
\begin{figure}[htp]
\centering
\includegraphics[angle=0,width=0.85\columnwidth]{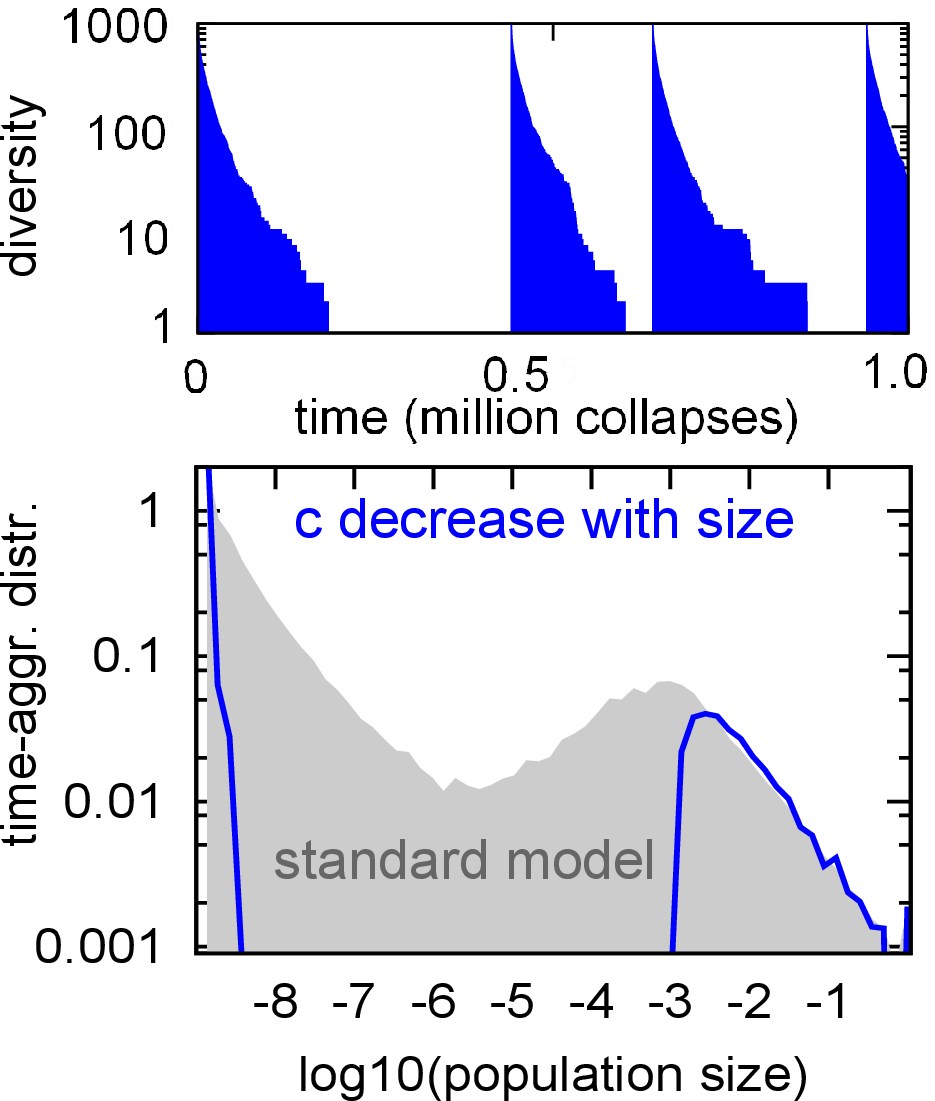}
\caption{\it {\bf ``Kill-the-loser (KtL) model''}in which the 
collapse probability declines with population size.  
The figure shows an $N=1000$, $\gamma=10^{-9}$ system 
simulated for $10^6$ collapse events.
The upper panel illustrates
the recurrent diversity waves, whereas the 
lower panel shows time-aggregated distributions, 
with the grey shaded area referring to our standard model.
}
\label{figS5}
\end{figure}

\item {\bf ``Fitness model''} with heterogeneous, 
species-specific growth rates and extinction probabilities.
Each species is assigned a growth rate $\Omega_i$ used when 
it repopulates the freed-up carrying capacity of the environment. 
It also has its own extinction probability $c_i$.
Both $\Omega_i$ and $c_i$ are 
logarithmically distributed in the interval between 
0.1 and 1. That is to say their $\log_{10}$ are 
uniformly distributed between $-1$ and $0$. 
At each time step we select one of 
$N$ populations, with probability
$\propto c_k$, this species goes extinct. It is immediately replaced
by a new species with the population 
$P_k \rightarrow \gamma=10^{-9}$, new growth rate 
$\Omega_i$, and extinction probability $c_i$.  
Subsequently all of the populations $i=1,2...N$ 
are rescaled proportional to their growth rates
\begin{equation}
P_i \rightarrow P_i+ (P_k-\gamma)  
\cdot \frac{\Omega_i P_i}{\sum_j \Omega_j P_j}
\nonumber
\end{equation}
to fill up the carrying capacity of the environment $\sum P_i=1$.
The upper panel in Fig. S6 shows that the
{\color{black} time-aggregated} population distribution in this model 
preserves its power-law tail, whereas
lower panel illustrates that in order for 
a species to reach substantial population size its 
fitness parameters need to be particularly favorable.
Indeed, populations larger than 
$1/N=0.001$ tend to have smaller than average extinction 
probabilities $c_i$, and larger than average growth rates $\Omega_i$.
\begin{figure}[htp]
\centering
\includegraphics[angle=0,width=0.85\columnwidth]{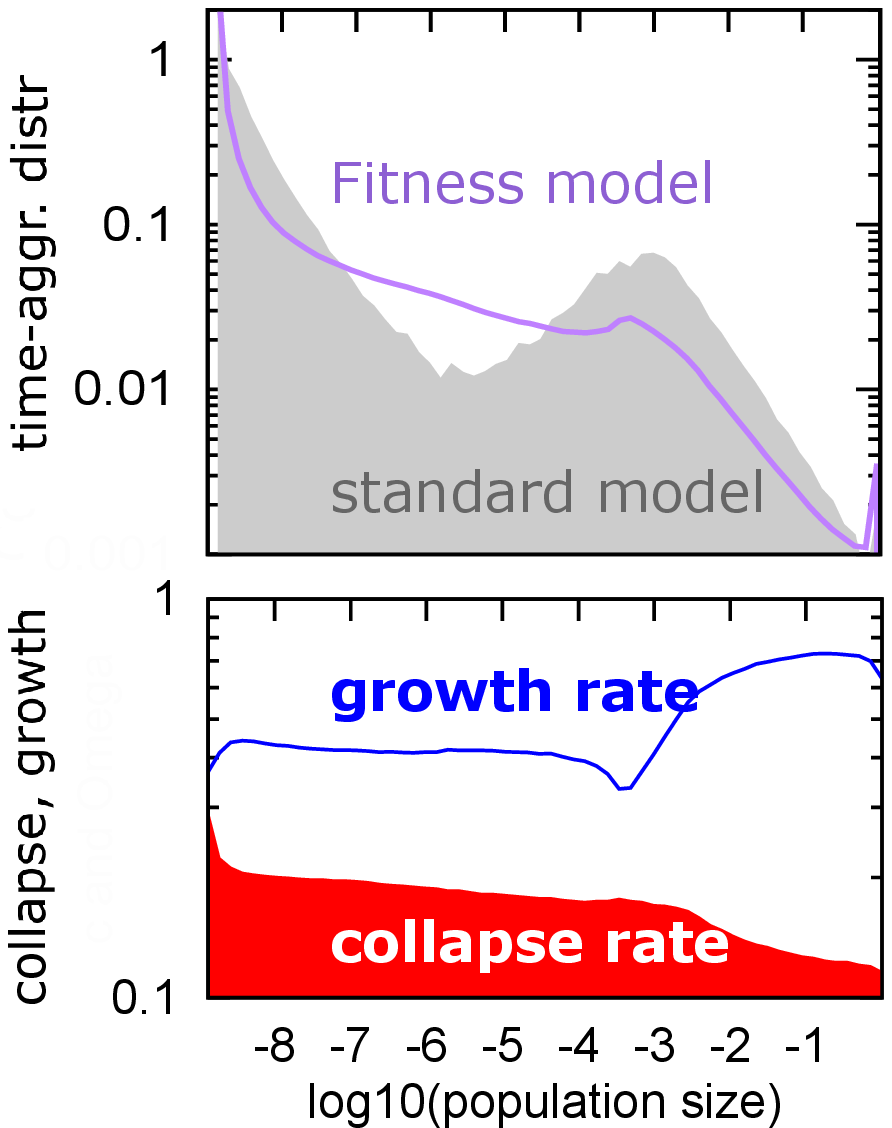}
\caption{\it {\bf ``Fitness-model''} with heterogeneous, 
species-specific growth rates and extinction probabilities.
Each species is assigned a growth rate $\Omega_i$ used when 
it repopulates the freed-up carrying capacity of the environment.
It also has its own collapse probability $c_i$.
Both $\Omega_i$ and $c_i$ are 
logarithmically distributed in the interval between 
0.1 and 1.
The purple curve in the upper panel shows
the {\color{black} time-aggregated} population distribution whereas 
the grey shaded area is that for the standard model where species' 
growth and collapse rates are all equal to each other. The
lower panel shows the average collapse probability and 
growth rate binned by the size of the population. 
he average growth rate $\langle \Omega_i \rangle$ 
(blue) and the average collapse 
probability $\langle c_i \rangle$ (red shaded area) of species 
binned by their collected at every time step. 
Both curves represent time-aggregated averages as 
individual populations change with time.
}
\label{figS6}
\end{figure}

\item {\bf ``Resilience model''} where heterogeneous, 
species-specific growth rates and survival ratios 
(population reduction following a collapse) are competing
with each other. Each species is assigned a growth rate 
$\Omega_i \in [0.1,1]$
and collapse size $\gamma_i \in [10^{-9},10^{-2}]$, both logarithmically
distributed (uniform distribution of the logarithm of the variable). 
At each time step we select one of 
$N$ populations, and collapse its population
$P_k \rightarrow \gamma_k \cdot P_k$. 
Note that unlike in previous versions we scale down the 
population proportional to its size and not proportional to
the carrying capacity of the environment. 
This reflects a different
interpretation away from our basic model,  where a collapse represents not an extinction of the species followed by the appearance of a new species at 
a fixed (very small) population size. In the new version of the model 
a collapse represents a sudden but proportionate reduction of a population e.g. 
due to species' phenotypic or genotypic bet-hedging. 
In this version of the model an 
extinction happens only if a very low population is reached, i.e. when 
$\gamma_k P_k<10^{-9}$. If this lower bound is reached 
the old species goes extinct and a new species 
with the initial population $P_k=10^{-9}$ is introduced. 
The new species is assigned new random values of 
$\Omega_k$ and $\gamma_k$. 
As in the previous model following each collapse 
all populations $i=1,2...N$ are rescaled 
proportional to their growth rates 
$P_i \rightarrow P_i+ (P_k-\max(\gamma_k P_k, 10^{-9}))  
\cdot \frac{\Omega_i P_i}{\sum_j \Omega_j P_j}
$ 
to fill up the carrying capacity of the environment: $\sum P_i=1$.
\begin{figure}[htp]
\centering
\includegraphics[angle=0,width=1\columnwidth]{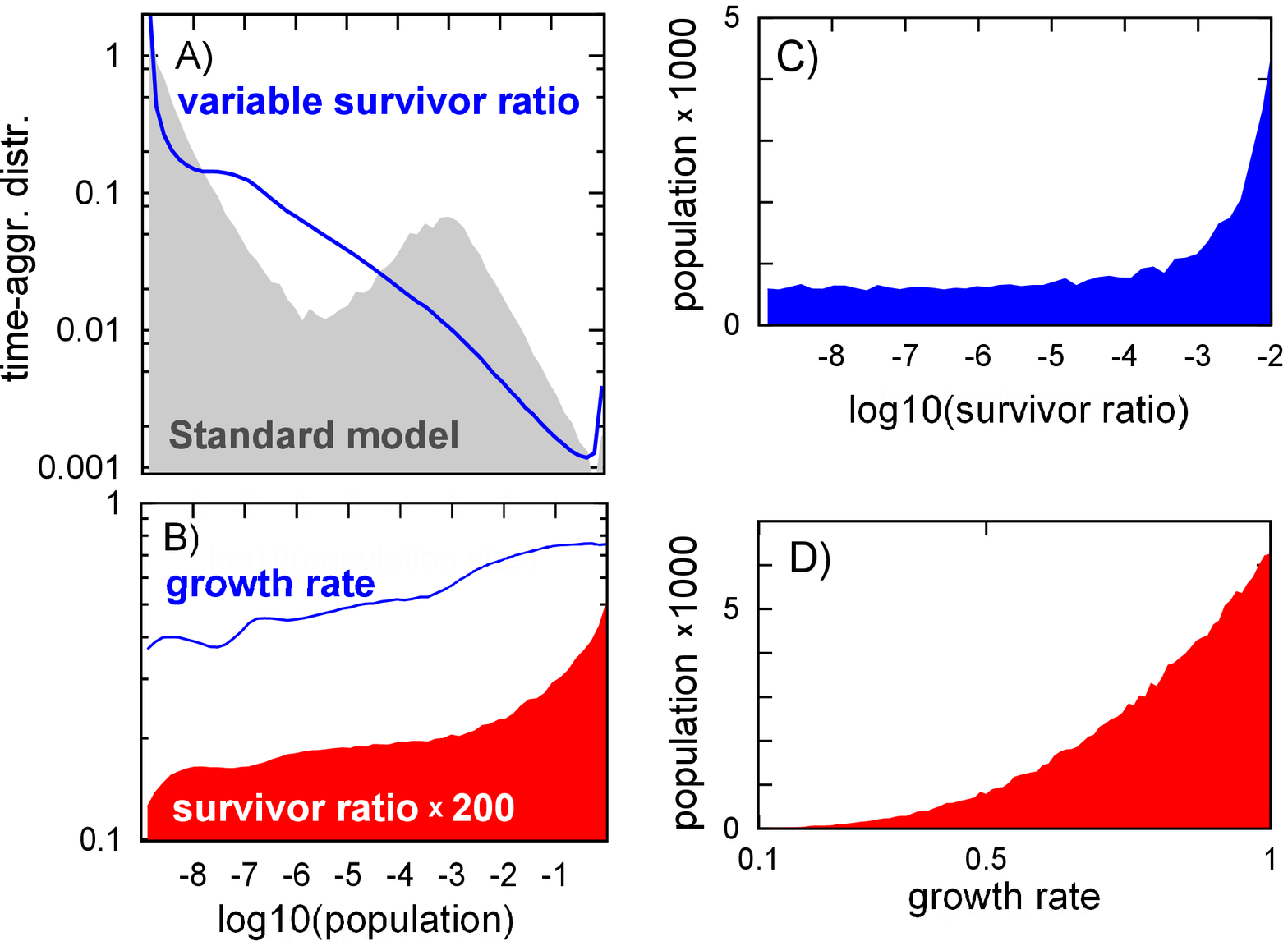}
\caption{\it {\bf ``Resilience model''} with heterogeneous, species-specific growth rates and survival ratios following a collapse. Each of the $N=1000$ 
species is assigned a growth rate 
$\Omega_i \in [0.1,1]$
and collapse size $\gamma_i \in [10^{-9},10^{-2}]$, both logarithmically
distributed. A) The blue curve shows the time-aggregated population distribution, whereas the grey area refers to that in our standard model from the main text. 
B) The average growth rate $\langle \Omega_i \rangle$ (blue) and the average survival ratio multiplied by 200 to 
have the same range in the plot $\langle \gamma_i \rangle$ (red shaded area)  as a binned by the population size collected at every time step. Both curves represent time-aggregated averages as individual populations change with time.
C) The average (arithmetic) population size as a function of species'
survivor ratio $\gamma_i$.
D) The average (arithmetic) population size as a function of species'
growth rate $\Omega_i$.
}
\label{figS7}
\end{figure}
\end{enumerate}

}

\end{document}